\documentclass[orivec]{llncs}
\usepackage[utf8]{inputenc}
\usepackage[T1]{fontenc}
\usepackage{graphicx}
\usepackage[table,xcdraw]{xcolor}
\usepackage{hyperref}
\usepackage{url}
\usepackage{longtable}
\usepackage[acronym]{glossaries}
\usepackage[square,numbers]{natbib}
\usepackage{tabularx}
\usepackage{enumitem} 
\usepackage{amsmath}

\usepackage{listings}

\usepackage{listings, xcolor}

\definecolor{verylightgray}{rgb}{.97,.97,.97}

\lstdefinelanguage{Solidity}{
	keywords=[1]{anonymous, assembly, assert, balance, break, call, callcode, case, catch, class, constant, continue, constructor, contract, debugger, default, delegatecall, delete, do, else, emit, event, experimental, export, external, false, finally, for, function, gas, if, implements, import, in, indexed, instanceof, interface, internal, is, length, library, log0, log1, log2, log3, log4, memory, modifier, new, payable, pragma, private, protected, public, pure, push, require, return, returns, revert, selfdestruct, send, solidity, storage, struct, suicide, super, switch, then, this, throw, transfer, true, try, typeof, using, value, view, while, with, addmod, ecrecover, keccak256, mulmod, ripemd160, sha256, sha3}, 
	keywordstyle=[1]\color{blue}\bfseries,
	keywords=[2]{address, bool, byte, bytes, bytes1, bytes2, bytes3, bytes4, bytes5, bytes6, bytes7, bytes8, bytes9, bytes10, bytes11, bytes12, bytes13, bytes14, bytes15, bytes16, bytes17, bytes18, bytes19, bytes20, bytes21, bytes22, bytes23, bytes24, bytes25, bytes26, bytes27, bytes28, bytes29, bytes30, bytes31, bytes32, enum, int, int8, int16, int24, int32, int40, int48, int56, int64, int72, int80, int88, int96, int104, int112, int120, int128, int136, int144, int152, int160, int168, int176, int184, int192, int200, int208, int216, int224, int232, int240, int248, int256, mapping, string, uint, uint8, uint16, uint24, uint32, uint40, uint48, uint56, uint64, uint72, uint80, uint88, uint96, uint104, uint112, uint120, uint128, uint136, uint144, uint152, uint160, uint168, uint176, uint184, uint192, uint200, uint208, uint216, uint224, uint232, uint240, uint248, uint256, var, void, ether, finney, szabo, wei, days, hours, minutes, seconds, weeks, years},	
	keywordstyle=[2]\color{teal}\bfseries,
	keywords=[3]{block, blockhash, coinbase, difficulty, gaslimit, number, timestamp, msg, data, gas, sender, sig, value, now, tx, gasprice, origin},	
	keywordstyle=[3]\color{violet}\bfseries,
	identifierstyle=\color{black},
	sensitive=true,
	comment=[l]{//},
	morecomment=[s]{/*}{*/},
	commentstyle=\color{gray}\ttfamily,
	stringstyle=\color{red}\ttfamily,
	morestring=[b]',
	morestring=[b]"
}

\usepackage{xcolor}

\definecolor{codegreen}{rgb}{0,0.6,0}
\definecolor{codegray}{rgb}{0.5,0.5,0.5}
\definecolor{codepurple}{rgb}{0.58,0,0.82}
\definecolor{backcolour}{rgb}{0.95,0.95,0.92}

\lstdefinestyle{mystyle}{
    backgroundcolor=\color{backcolour},   
    commentstyle=\color{codegreen},
    keywordstyle=\color{magenta},
    numberstyle=\tiny\color{codegray},
    stringstyle=\color{codepurple},
    basicstyle=\ttfamily\footnotesize,
    breakatwhitespace=false,         
    breaklines=true,                 
    captionpos=b,                    
    numbers=left,                    
    numbersep=5pt,                  
    showspaces=false,                
    showstringspaces=false,
    showtabs=false,                  
    tabsize=2
}
\usepackage{xcolor}
\hypersetup{
    colorlinks,
    linkcolor={blue!80!black},
    citecolor={blue!50!black},
    urlcolor={black!80!black}
    }
\usepackage{appendix}
\usepackage[colorinlistoftodos]{todonotes}
  
\usepackage{xspace}

\begin{document}

\title{Safeguarding Physical Sneaker Sale\\Through a Decentralized Medium}

\author{
Marwan Zeggari\thanks{marwan.zeggari@lyzis.tech}\inst{1}
\ \ \ Aydin Abadi\thanks{aydin.abadi@ucl.ac.uk}\inst{2}\\
Renaud Lambiotte\thanks{lambiotte@maths.ox.ac.uk}\inst{3}
\  \ Mohamad Kassab\thanks{muk36@psu.edu}\inst{4}}
\institute{Lyzis Labs, USA \and University College London, UK \and University of Oxford, UK \and Pennsylvania State University, USA}

\date{\today}
\maketitle{}

\begin{abstract}
Sneakers were designated as the most counterfeited fashion item online, with three times more risk in a trade than any other fashion purchase. As the market expands, the current sneaker scene displays several vulnerabilities and trust flaws, mostly related to the legitimacy of assets or actors. In this paper, we investigate various blockchain-based mechanisms to address these large-scale trust issues. We argue that (i) pre-certified and tracked assets through the use of non-fungible tokens can ensure the genuine nature of an asset and authenticate its owner more effectively during peer-to-peer trading across a marketplace; (ii) a game-theoretic-based system with economic incentives for participating users can greatly reduce the rate of online fraud and address missed delivery deadlines; (iii) a decentralized dispute resolution system biased in favour of an honest party can solve potential conflicts more reliably.
\end{abstract}

\section{Introduction}

Digital technologies are contributing to an ever-changing retail space and are reshaping the way purchases are made online.
Many markets happen to be the subject of growing consumer focus, partly due to their stimulated interests in social media and partial speculation being an appealing economic trend \cite{spec}. 
The sneaker resale industry reflects all the properties of this form of concentrated demand market, where items sell for more than the original price. This market demand is mainly fueled by strong exposure and commercial strategies (e.g., celebrities and designers' involvement, limited quantity items, individualistic cultures) \cite{sec}.
However, a secondary market\footnote{A secondary market is considered to be a market where parties sell and buy assets they already own.} with speculative inclinations such as the sneaker sector can result in numerous constraints from both a seller's and a buyer's perspective. Several concerns of trust and legitimacy have been expressed in relation to the traded products as well as the participants involved \cite{tr}.
Product counterfeiting, payment scams, and phishing are some of the most common concerns and obstacles when conducting a peer-to-peer (P2P) sneaker sale.
Within this scope, several emerging technologies, notably blockchain technology and its by-products, can be qualified as solutions to these transactional complications, allowing overall a more reliable exchange space.
The constant shift in markets and personal or economic behaviour of consumers can drive the adoption of these technological solutions as long as there are constant benefits to these actors involved.

This paper focuses on the use of a P2P marketplace based on blockchain technology, i.e., Lyzis Marketplace \cite{lyz}, which helps to address some of the current limitations experienced in the commercial landscape of sneaker trading between peers.
Our findings essentially suggest that (i) leveraging the use of a distributed registry and underlying Non-Fungible Tokens (NFTs) to identify and authenticate physical products is more effective than the means currently used, (ii) relying on a decentralized architecture using several defined mechanisms allows to bring more trust between the peers involved, and (iii) a potential involvement of these actors in the process of governance at some level can lead to a more assured overall equilibrium and fairness than with the means of selling and buying sneakers traditionally used.

This paper is organized as follows.
In Section \ref{Market}, we provide a general insight into the landscape of the sneaker retail market as well as the challenges  end users may be often experiencing.
In Section \ref{Platform}, we introduce our solution, Lyzis Marketplace, and describe its initial technical architecture and application to the defined sneaker industry.
Finally in Section \ref{Solution}, we describe all the mechanisms used within our platform to address the perceived problems of the target market.

\subsection{Related Works}

Many studies, such as \cite{pavel, batra, white}, deal with marketplaces for NFTs trading but do not cover the resale market of physical sneakers.
Only a few papers address or refer to blockchain and/or NFT-based systems in the field of sneakers, such as in \cite{NFTsneakmarket, Nextgen, pharma}.
Other works such as \cite{zhu, choi, raditya, ma, cassidy} have specifically studied the dynamics of the sneaker resale market, including auction models, consumer perception, sales model optimization or price discovery but do not address market-related issues and lack a decentralization-based approach.
Further surveys such as \cite{alex, umer, emma, joy, raman} focus more broadly on the luxury market as a whole, demonstrating the added value of using NFTs for brands to improve customer loyalty and shoppers in terms of ownership identification and transparency.
Being however fairly limited and covering for some only the architectural side of the implementation of a sneaker marketplace, we mainly suggest in this paper to approach this market by extending the features of Lyzis Marketplace, a blockchain-based and trustless trading platform, incorporating NFTs and leveraging game theory for peer-to-peer exchanges of physical assets introduced in a previous work \cite{lyz}. 

\section{Background: The Sneakers Sales Market}\label{Market}

Collaborative fashion consumption\footnote{Defined in \cite{collab} as a consumption trend in which consumers, instead of buying new fashion products, have access to already existing garments either through alternative opportunities to acquire individual ownership (gifting, swapping, or second-hand), or through usage options for fashion products owned by other (sharing, lending, renting, or leasing).} makes up a huge industry, and the sneaker business\footnote{According to \cite{sneakmarket}, the secondary sneaker market originally consists of trading or selling rare and limited edition sneakers and streetwear through auctions on online platforms or through reorganized social media groups run by consumers.} draws a sizable audience. 
In fact, the state of the secondary sneaker market has rarely been healthier with an overall size of \$6 billion in 2022, a primary market of \$100 billion, and a projected secondary to primary market size in 2025 ranging from 15 to 25\%, growing steadily \cite{datastockx}.
Assets in the form of sneakers are highly prized by consumers, whether for collection, price appreciation, and/or investment purposes \cite{assetsneak}. 
Sneakers have become a sign of ``absolute prestige'', which was driven by the increasing collaborations between casual culture and celebrity endorsements.
These were observed to have an impact on prices and quantities available, resulting in high expectations \cite{2ndmarket}. 
Studies also reveal that a user's acceptance and intention to eventually patronize a collaborative fashion consumption platform depends on ownership and trust \cite{distrust}, two features that we strive to embed within our solution in a secure and decentralized way upfront. Therein lays the main basis for the initial application of the Lyzis Marketplace to this specific type of market.

Today, the secondary sneaker market has witnessed the growth of many intermediaries (i.e., professional online marketplaces), most notably the largest, StockX \cite{StockX} and GOAT \cite{GOAT}. 
These centralized intermediaries - taking a share of the sale of each pair of sneakers - target a rather digital-native audience, mainly composed of relatively young users. These young consumers are usually keen to convey a statement about their status or taste in footwear to other people, generally publicly whether on social networks or around campuses \cite{collab}. 
This is the key driver\footnote{Indeed, popular culture and social media are frequently credited as significant contributors to the interest in the sneaker submarket \cite{sneakmarket, expen, driver}.} behind this sector's hyper-growth\footnote{According to \cite{market}, the sneaker market has grown 46\% globally since 2017, and shoes such as Nike's Air Jordans are considered a fashion statement rather than a basketball shoe. Globally, the sneaker market is expected to grow between 2019 and 2024 at a Compound Annual Growth Rate (CAGR) of more than 7\% and reach \$88 billion by 2024.}. However and in spite of lightning growth, many issues persist in the retail sneakers scene. This is particularly the case when it comes to the authenticity of an asset (e.g. counterfeit product or certificate of authenticity) and/or the legitimacy of an actor (e.g. payment or phishing scams, missed deadlines). The next section elaborates on the types of recurring challenges faced by users of centralized platforms to buy and sell sneakers.

\subsection{Established Issues}\label{issues}

In  this section, we discuss several established issues in the online sneaker market. The issues have been reported by the users of StockX and GOAT platforms. See Appendix \ref{AppendixA} for further insight into the figures retrieved.

\subsubsection{Counterfeit assets:} The sale of counterfeit or fake shoes is one of the most popular sneaker resale frauds. In this case, a seller provides a buyer with a substitute, low-quality pair by assuring him of its veracity, usually with an accompanying authentication certificate being falsified. To illustrate the scale with real-world figures, it can be stated that in 2022 the counterfeit sneaker market is worth \$450 billion \cite{WashingtonPost}, more than 5 times the value of the legitimate market estimated to be around \$85.54 billion \cite{runrepeat}.

\subsubsection{Payment scams:} Another form of scam that may be experienced in the sneaker resale market is the payment scam, where a seller asks for upfront payment but does not deliver the sneakers or a buyer gets the pair without paying through intermediate methods. In that situation, the refunding process is frequently drawn out and takes several weeks, which is too long for the customers. For example, a user sold a pair of sneakers in 2022 and received a legitimate email from the buyer protection program stating that the money would be released after sending the item's tracking number. Since the email in question was not issued by the relevant entity (Paypal), the seller has no way of retrieving his funds \cite{Paypalscam}. Such a scam is known as a ``purchase scam'' (as a variant of ``Push Payment Fraud'') in the UK; according to ``UK Finance'', the total value of purchase scams in the UK was  over $£$210m, only in the first half of 2021 \cite{2021-Half-Year-Fraud-Update}.

\subsubsection{Phishing scams:}  In some cases, scammers may use fake emails or websites to try to trick buyers into providing personal or financial information or to trick sellers into providing access to their accounts. For instance, a new phishing scam concerning Adidas models has been detected, carried out in 2019 and then in 2021. It offers free shoes and money via email using the Adidas brand name. The messages claim that Adidas is celebrating its 93rd birthday and is offering 3,000 lucky customers a free pair of Adidas sneakers and a free subscription of \$50 per month \cite{adidasscam}. Users must then send their personal data to be eligible.

\subsubsection{Poor customer service:} Bad customer service poses a real threat to online marketplaces specialised in sneakers. Customers usually complain about bad customer service when \textit{(i)} one issue arises during the order and they are unable to get assistance in the process, \textit{(ii)} customer support takes too long to address their issues and \textit{(iii)} the final responses provided is not satisfactory without being able to reprocess the case. This issue is a major contributor to the seller and buyer frustration, with an average of 67\% of online customers having become ``serial switchers'', customers who are willing to switch brands because of a poor customer experience \cite{customerservice}.

\subsubsection{Ineffective cancellation policy:} The lack of a solid and generous cancellation policy often results in the client paying for a fake pair and not being able to be refunded. This issue arises especially with the presence of bots during large drops\footnote{A drop is the limited sale of a special type of sneakers that are either scarce in quantity (due to limited production) or scarce in availability (due to a temporary purchase window). }. If a buyer manages to order a pair, it may no longer be eligible for a cancellation or refund \cite{adidas}.

\subsubsection{Not accepting returns:} This issue is inextricably linked to the shop's policy and is therefore correlated to the ineffective cancellation policy. However, this issue persists for buyers receiving a different item than the one displayed. Sneaker resale websites usually have a very strict policy towards customers.

\subsubsection{Hidden fees:} This represents a huge challenge when dealing with third-party sneakers being delivered from
 foreign countries. Hidden fees mostly represent delivery fees accounting for 100\%+ of the original order. By way of example, we notice that in April 2023, a lawsuit alleges that certain prices posted on StockX's website \cite{StockX} hide various fees from customers, in violation of the Quebec Consumer Protection Act \cite{lawsuit}. This case is then based on the platform's pricing systems, which are not always fully transparent.

\subsubsection{Delivery time issues:} In this situation, delivery times are not met and it often takes several weeks/months to be delivered. This process erodes customer confidence and trust in the online marketplace. Here, the origin and sender of the pairs often play a determining role on the delivery time, with sneakers usually sent from foreign countries by legal entities rather than peer users acting on their behalf.

\subsubsection{Low User Experience (UX):} The UX is extremely important when it comes to the purchase of high-priced B2C items. This minor but important issue leads customers to sometimes buy several pairs instead of one.

\subsection{Our Focus}

Through our work, we address and resolve, within an initial framework, many issues surrounding the sneaker resale market as identified in the previous Section \ref{issues} including counterfeit assets, payment scams, poor customer service, inefficient cancellation policies, unaccepted returns, hidden fees, and missed delivery times. 
Cases of phishing and low UX will be addressed in future work.

\section{Lyzis Marketplace: A Blockchain-based Middleware}\label{Platform}

By design, Lyzis Marketplace - initially introduced in \cite{lyz} and for which we extend the features to fit the sneaker market - allows two mutually distrustful parties to trade any non-digital asset by rolling out a smart contract over a blockchain to act as an escrow.
The necessary components are embedded to ensure that within the platform, an actor's (i.e., seller's or buyer's) honest strategy is safe to the fullest extent - in a strong game-theoretic sense (see Section \ref{gametheory}) - if the arbiter is biased in favour of the honest parties (see Section \ref{conflict}).
This mainly leads to a significant reduction of potential conflicts online, and to a more reliable and secure trading space.

\subsection{Initial Structure}

A blockchain-based venture can display extensive features attributable to data security, governance, and user privacy, as well as a significant reduction in intermediate costs for completing an exchange, contract, follow-up, and more \cite{fincontr, trackmanagehealth, vote}. This may help address the end-users commonly held trust issue with centralized solutions and the limited transparency of their rules and conditions, letting actors rely entirely on distributed networks. The latter is the main driver behind the design of Lyzis Marketplace on this kind of system.
The proposed decentralized architecture for the operation of our platform, the built-in modules and their interfaces are illustrated in Fig.\ref{Fig2}, Appendix \ref{AppendixB}. 
Here, the platform acts as a direct intermediary through smart contracts between the peers involved without needing an associated central entity.

\subsection{Decentralized Litigation Management}\label{conflict}

To settle Lyzis Marketplace-related disputes, we rely on the Kleros justice protocol \cite{lesaege2021long}.
Kleros is a blockchain-based dispute resolution layer that provides fast, secure and cost-effective arbitration for all online disputes (an opt-in court system). Kleros acts in Lyzis Marketplace as a decentralized third party to arbitrate disputes during the exchange period and afterwards. It essentially relies on game theoretic incentives to have jurors rule cases correctly \cite{kleros}. 
Reference \cite{lyz} to view how exactly Kleros is used within Lyzis Marketplace.

\section{Our Solutions}\label{Solution}

To address the challenges identified in Section \ref{issues}, we use several mechanisms and tools/techniques within our framework.
We basically suggest: (i) the use of Non-Fungible Tokens (NFTs) to guarantee authenticity and track an asset's ownership, (ii) a game theory-based distributed architecture to ensure increased transparency of P2P transactions, and provide economic and governance incentives to act honestly toward a social optimum, and (iii) a decentralized dispute management system coupled with an insurance model biased in favour of honest parties.

\subsection{An Overview of our Solution }

Below, we present an overview of our proposed mechanisms when   
 a seller and buyer interact with each other.

\begin{enumerate}
    \item A user submits a pair of sneakers for sale on the platform by entering the corresponding information and pictures. (see Appendix \ref{hybridassets} for details).
    \item Once the pair is authenticated by an external party (e.g., CheckCheck \cite{legit}), an NFT is generated on-chain, having the data incorporated by the seller hidden in its metadata (see Section \ref{C2} for details). This NFT, linked to the pair of sneakers and its attributes, guarantees the authenticity of the product and the accuracy of its certificate and allows to  track its ownership.
    \item The pair of sneakers is available online and visible to all potential buyers.
    \item A buyer is interested in the sneakers. He discusses with the buyer and agrees on purchasing and shipping terms. The buyer then deposits the required amount, in tokens/cryptocurrencies, into the smart contract dedicated to the exchange from a Lyzis Marketplace interface. The seller then initiates the shipment within the given timeframe. Both have an economic incentive to act honestly during the whole exchange, i.e., they are eligible to receive governance tokens (LZSP\footnote{The LZSP governance token is a token generated to give voting rights (ratio of 1 token = 1 vote) in the Lyzis Decentralized Autonomous Organization (DAO). The LZSP is redeemable for LZS, the core utility token for value transfer.}) when they act honestly in pre-defined actions, e.g., shipment deadlines met, information provided on the shipment's progress, receipt confirmation, and reviews assigned. They may also be penalized for choices that are considered dishonest.
    \item Once the buyer has received the pair of sneakers and is satisfied, the NFT previously generated and linked to the pair has its owner attribute updated (see Section \ref{hybridassets} for details).
    \item The trade is complete. The NFT is now in the possession of the buyer and is still linked to the same pair he received. The seller has obtained his tokens, representing the value of the pair. In the event that a problem arises during the interaction between the two users and/or the outcome of the exchange is considered unfair by any of the parties, (e.g., the item is not as agreed or in bad condition, the item is eventually fake but the certificate of authenticity is true or inversely or both are fake) then the wronged user can first appeal to an online decentralized dispute management tribunal\footnote{To settle Lyzis Marketplace related litigation cases, we propose to rely on the use of the Kleros justice protocol \cite{lesaege2021long}. Kleros is a blockchain dispute resolution layer that provides fast, secure and cost-effective arbitration for all online disputes (an opt-in court system). Here, a random jury is then selected from around the world, all with an economic incentive to be biased in favour of the honest party, to give a verdict on the case submitted by the users. For more details on how the Kleros protocol is applied within Lyzis Marketplace, refer to \cite{lyz}.}, and be eligible for a refund through a decentralized insurance protocol (see Section \ref{insurance} for details) as long as he can prove his case.
\end{enumerate}

\begin{figure}[h!]
\centerline{\includegraphics[width=1\textwidth]{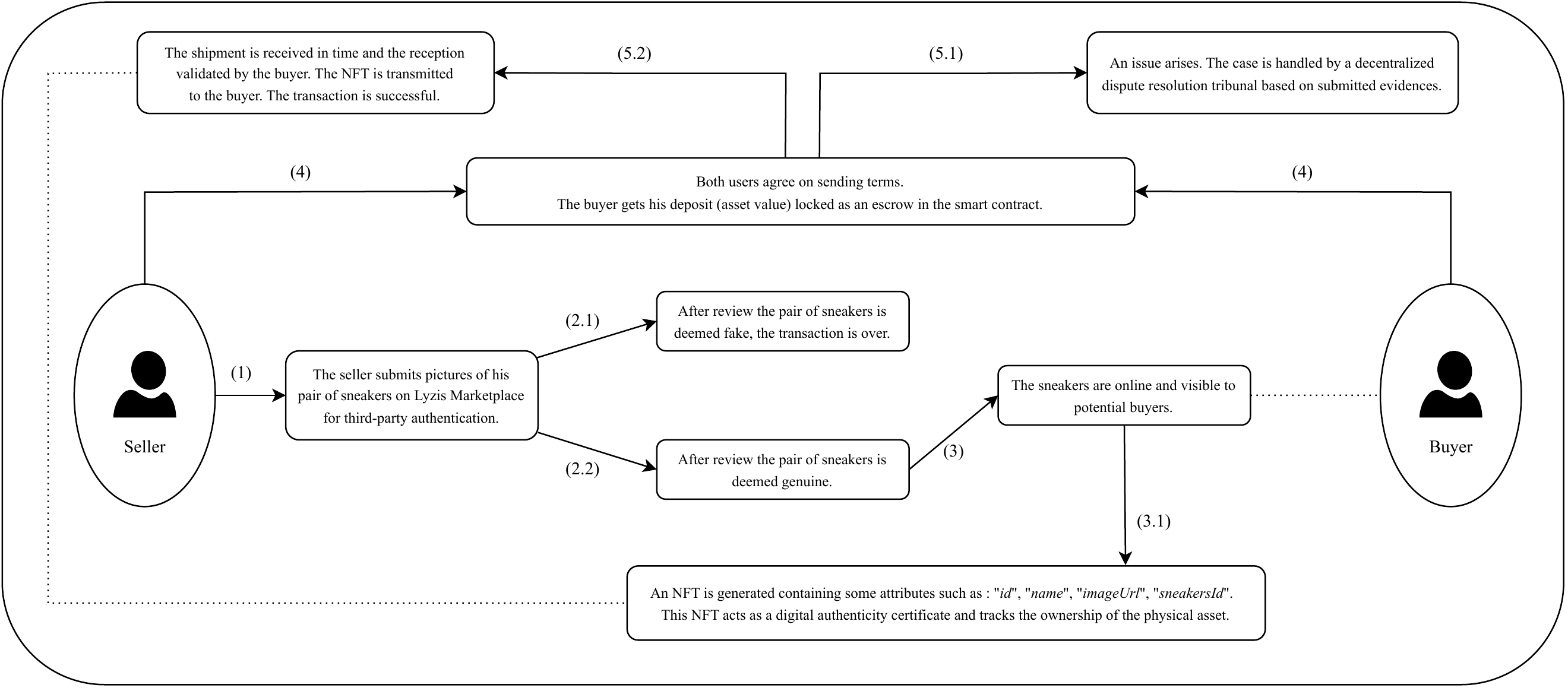}}
\caption{Sneakers trading workflow between a buyer and a seller through Lyzis Marketplace.}
\label{Fig0}
\end{figure}

Fig.\ref{Fig0} summarizes the above trade description and the mechanisms in operation during an interaction between two users seller/buyer.

\section{Details on Implemented Features}

In this section, the features, mechanisms, and tools implemented within Lyzis Marketplace as solutions are covered in detail.

\subsection{Hybrid Assets: A Guarantee of Property Rights Using Non-Fungible Tokens (NFTs)}\label{hybridassets}

NFTs are non-fungible tokens that we propose to use to represent the ownership of single assets, in our application case the physical sneakers. The ownership of an asset is immutable and secured on the Ethereum blockchain. Nobody is able to alter the ownership record except in the case of a property transfer\footnote{Since each record is linked to the previous and subsequent records on a distributed ledger, attackers would have to modify the entire chain to change a single record - making it difficult to modify.}. Several distinct attributes\footnote{Term extracted from the RFC2459 \cite{attr}. With the ERC-721 token standard, \textit{attributes} are coded into NFTs on-chain metadata and it refers to all on-chain data related to the NFT. This standard allows an application to run the Ethereum standard API for NFTs in a smart contract.} are determined when creating an NFT.
The following are the characteristics of our case study when submitting a pair of sneakers for sale\footnote{We assume that selling users have valid and functional accounts linked to a wallet, custodial or not, that matches the criteria defined in \cite{lyz}.}.

\begin{itemize}
    \item[-]{The 3 primary attributes of the NFT are generated, namely \textit{``id''}, \textit{``name''} and \textit{``imageUrl''}.}
    
    \item[-]{Additional attributes are subsequently added, namely \textit{``sneakerId''}, \textit{``location''}, \textit{``proofOfOwnership''}, and \textit{``transferOfOwnership''}.}

    \begin{itemize}

    \ 
    
    \item{\textit{sneakerId}: This attribute is the unique identifier for the physical pair of sneakers. It can be the serial number of the sneaker\footnote{For instance, in the case of Nike sneakers, each pair consists of 9 numbers or letters, the first 6 of which are the shoe code and the next 3 the colour code.}, a unique number that we generate, the number that appears on the sneaker's certificate of authenticity, or a combination of them.}

    \ 
    
    \item{\textit{location}: This attribute represents details on the physical storage of the sneakers, i.e. owner, location, store/brand, etc.}

    \ 
    
    \item{\textit{proofOfOwnership}: This attribute represents the proof of ownership of the physical sneakers. It can be a certificate of authenticity picture coupled with a sneaker picture.}

    \ 
    
    \item{\textit{transferOfOwnership}: This attribute is a mechanism to transfer the ownership of the NFT and the physical sneakers included in the smart contract. It may involve updating the ``owner'' attribute and recording the transfer in the Ethereum blockchain.}
\end{itemize}
\end{itemize}

These attributes included in the metadata recorded on-chain provide a link to the physical asset, i.e., the pair of sneakers, hence the term hybrid asset. An added redirection exists - for instance with the associated serial number - between the online representation of the asset and the physical product.

An example of a functional NFT smart contract, in the Solidity language - deployable and linkable to a physical pair of sneaker - which includes the attributes and features illustrated above is shown in List.\ref{List1}, Appendix \ref{AppendixB}.

To safeguard users' privacy, data considered as sensitive may be encrypted to be divulged to authorized parties only (see Appendix \ref{C2} for more details).

\subsubsection{Preventive mechanisms}

Several mechanisms must be used to prevent constraints when running hybrid assets \cite{prox}. Among these potential constraints, in our situation where a hybrid asset links an NFT to a pair of physical sneakers, the following cases can be considered.

\begin{itemize}

\item\textbf{Strong identification of asset and/or asset providers:} In order to initially guarantee maximum authenticity of the provided assets and their certificates set in NFT, we propose a verification step performed through an external entity in charge of authenticating the asset\footnote{The asset pictures for authentication purposes must then be submitted live from the dedicated interface by the seller. This adds security in case a dispute arises later on, allowing the court in charge of the dispute to have a reliable support on which to base its decision, i.e., certified photos dated before the shipment.}, i.e., CheckCheck \cite{legit}, coupled with a KYC (Know Your Customer) process to identify the asset provider\footnote{It may also be possible to rely on decentralized arbitration protocols, e.g., Kleros protocol \cite{lesaege2021long}. These alternatives are to be explored during implementation.}.
Additionally, a user is subject to penalties from the governance committee if an asset turns out to be non-authentic afterwards (see Section \ref{governance}). If a user has acquired a pre-certified non-authentic asset, then they are eligible for a refund as long as they can prove their cases (see Section \ref{insurance}).

\item\textbf{Prevention of simultaneous on-chain double-spending:} To provide assurance to a potential buyer that the same pair of off-chain sneakers is not transformed into multiple NFTs on different blockchain networks, only specific blockchains must be supported and considered when generating the NFT. The relevant blockchain must then be explicitly referred to when buying a pair of sneakers and the buyer must always know which blockchain is being used.

\item\textbf{Standard mechanism for hybrid assets:} In our hybrid asset case, i.e., NFT bound to a pair of physical sneakers, it is possible to rely on oracles for a mandatory attributes change (e.g., location) and to have a perfect synchronization between off-chain to on-chain state changes.

\end{itemize}

\subsubsection{The Asset Proxy NFT: A computational model to ensure consistency between on-chain assets and the metadata of corresponding off-chain physical sneakers.}
\begin{figure}[h!]
\centerline{\includegraphics[width=1\textwidth]{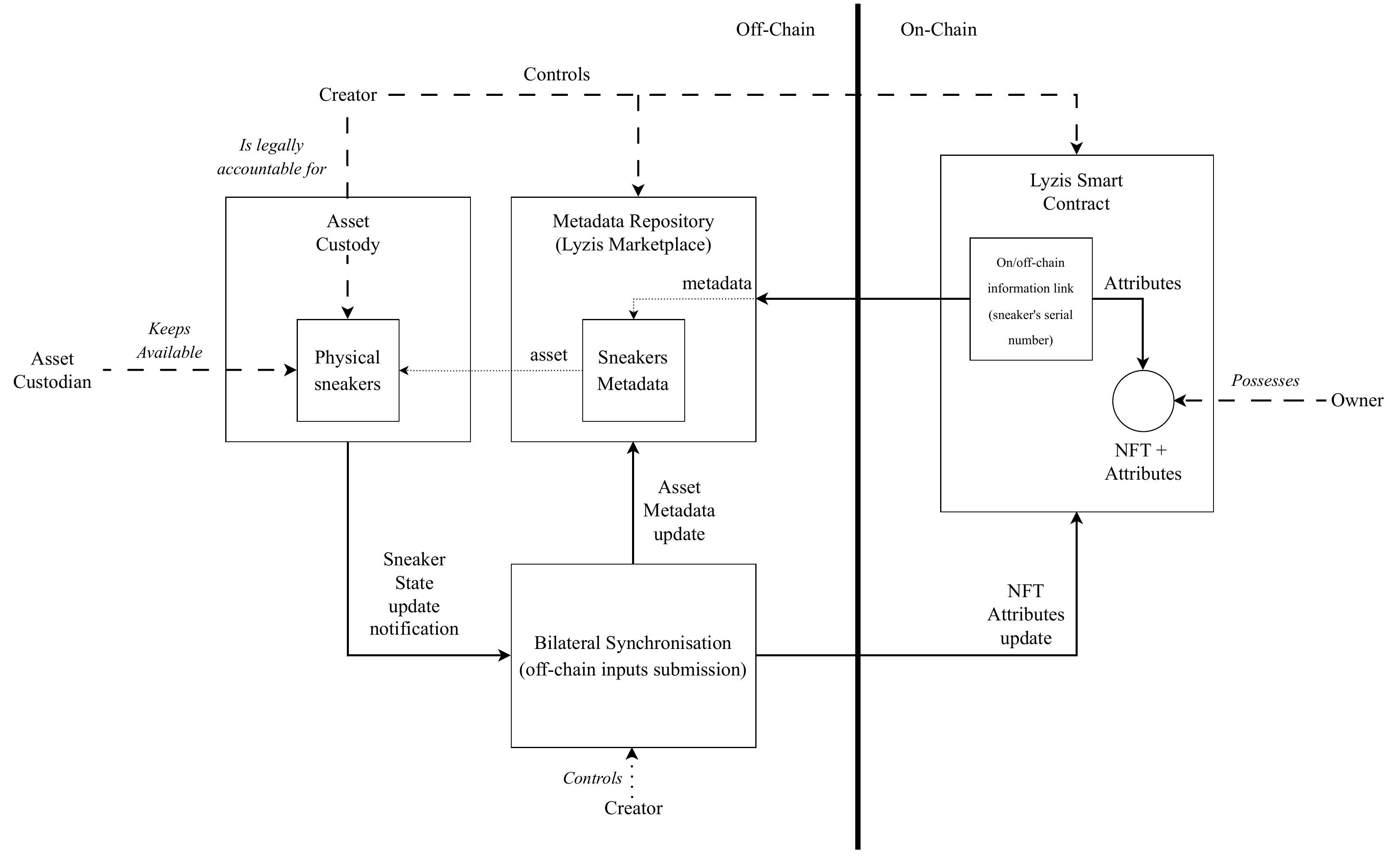}}
\caption{Asset Proxy NFT architecture \cite{prox}  applied to the NFT exchange and matching model to physical sneaker pairs.}
\label{Fig1}
\end{figure}

A recurring challenge in the field of NFTs is to maintain the greatest possible consistency between the token (on-chain) and its corresponding metadata (off-chain).
The Asset Proxy NFT's role is then to provide the ability to achieve technical state consistency between off-chain assets and NFTs as on-chain representations of those assets \cite{prox}. The generated NFT is then totally bound to the state and changes of its assigned sneaker peer.
Based on the architecture of the NFT Asset Proxy provided in \cite{prox}, we are able to illustrate in Fig.\ref{Fig1} the components and interfaces of the NFT Asset Proxy applied to Lyzis Marketplace.

We assume that a selling user (i.e., seller) acts as Creator, Asset Custodian and Owner (before a transaction). Below is a basic case of how a sneaker sale on the platform works and how it generates an NFT associated with it.

\begin{enumerate}
    \item The selling user (i.e., creator, asset custodian and/or owner) is in possession of a pair of sneakers and the related authenticity certificate (\textit{Asset Custody service initialized}).
    
    \item The necessary configuration components are added by the same selling user through a Lyzis Marketplace interface, including information about the pair and/or other necessary attributes (pictures, details, etc.) (\textit{Metadata Repository service initialized}).
    
    \item Finally, and after due diligence on the asset and the supplier\footnote{Due diligence is conducted on the asset by a third party audit, and on the supplier by a KYC linked to the platform.}, the NFT smart contract linked to the sneakers pair is deployed (see List.\ref{List1}, Appendix \ref{AppendixB}.). 
    On/off-chain information link (e.g., sneaker's serial number) is explicitly implemented by the selling user and is immutable\footnote{Note that no complex bilateral synchronization mechanism is needed as no changes are possible in the NFT attributes once recorded (i.e., fixed serial number) - a simple redirect link/interface may be used. In case a major change is absolutely required in the NFT attributes (e.g., location, owner), it is possible to rely on the use of Oracle. This bilateral synchronization component, controlled by the asset creator, ensures full consistency between the sneakers state, its metadata, and the attributes of the NFT token \cite{prox}.}, i.e., no more reconfiguration possible.
\end{enumerate}

Appendix \ref{AppendixC} expands on additional details and requirements for the use of NFTs as a physical asset tracker.

\subsection{A Game-theory Based System to Earn by Trading: Valuing Honesty}\label{gametheory}

Our system is based on the assumption that, in order to foster overall efficiency, individual market participants' behavior must have an economic incentive to be honest rather than dishonest. Thus and in consideration of honest and beneficial behavior, users are rewarded by trading on the platform.

The model that we implement allows the use of incentives for good behavior and disincentives for bad behavior. This general system  is initially consisting of three components: a LSZP payoff (governance token) for defined honest actions, a reputation system with measurement points and a minimum stacking amount to keep a new account validated and active on the platform (refer to \cite{lyz} for more details on the mechanisms involved).
This practice is supported by a mathematical field called Game Theory, which models strategic choices and interactions between rational decision-makers \cite{commnet}. 

In a game, each decision maker (as a player) chooses his strategy to maximize his utility given the possible strategies of the other players. In other terms, if a user knows that he has too much to lose by acting dishonestly and much to gain by acting honestly, he will rationally prefer to act honestly or not to act at all.
To describe and model a situation representing a game (i.e., decisional interaction situation), the following elements must be considered.
The set of players (or decision-makers) and their possible actions, the rules of the game specifying, in particular, the order in which the players play and when the game ends, and the possible outcome of the game for each player and its implication in terms of ``payoff function'' (to do so, we need to know players' preferences).

Then, we can briefly formalize the strategies that can be adopted within Lyzis Marketplace and the associated results (payoffs) in the form of a simultaneous and non-cooperative\footnote{Term widely used in Game Theory to express a situation in which we assume that the players are completely independent of their decisions at the time they make their selections. Also, the sequential nature of this game means that $i_{1}$ and $i_{2}$ decide at the same time on the strategy they are going to adopt (when placing the order and validating the sale). } game with 2 players where:

\begin{center}
    $i_{1}$ = The buyer buying an item from the seller $i_{2}$;
    
    $i_{2}$ = The seller selling an item to the buyer $i_{1}$.
\end{center}

The possible strategies adopted by the players $i_{1}$ and $i_{2}$ are then represented\footnote{The model can be generalized to include multiple types of fraudulent behavior.} by:

\begin{center}
    $s_{1}$ = Player behaves honestly;
    $s_{2}$ = Player behaves dishonestly.
\end{center}

Moreover, in this game, each player gets a \textit{payoff} that depends on its own strategy and the strategy of its interlocutor. Tab.\ref{tab2}, Appendix \ref{AppendixB}, summarizes these payoffs. 
Considering the variables and functions defined above, we obtain Tab.\ref{tab1} which models the game theory applied within Lyzis Marketplace for a sneaker sale.
We provide next a brief summary of the outcomes for the specific strategies observed (refer to \cite{lyz} to see the full development).

\begin{table}[h!]
\renewcommand{\arraystretch}{1.2}
\setlength{\tabcolsep}{0.12cm}
\begin{center}
\begin{tabular}{|ccc|}

\hline
$i_{1}/i_{2}$  & $i_{2}\sim s1$                                                                                                                                            & $i_{2}\sim s2$                                                                                                                                                            \\

$i_{1}\sim s1$ & \begin{tabular}[c]{@{}c@{}}$vo_{1},\alpha\sim LZSP_{1}, \beta\sim S_{3}, (\delta\sim R_{1}); $\\$ vo_{1}, \alpha\sim LZSP_{1}, \beta\sim S_{1}, (\delta\sim R_{1})$\end{tabular} & \begin{tabular}[c]{@{}c@{}}$vo_{1}, \alpha\sim LZSP_{2}, \beta\sim S_{4};$\\ $  vo_{2}, \alpha\sim LZSP_{2}, \beta\sim S_{2}, (\delta\sim R_{2})$\end{tabular} \\

$i_{1}\sim s2$ & \begin{tabular}[c]{@{}c@{}}$vo_{2}, \alpha\sim LZSP_{2}, \beta\sim S_{3}, (\delta\sim R_{2});$\\ $vo_{1}, \alpha\sim LZSP_{2}, \beta\sim S_{1}$\end{tabular}  & \begin{tabular}[c]{@{}c@{}}$vo_{1}, \alpha\sim LZSP_{2}, \beta\sim S_{3};$ \\ $ vo_{1}, \alpha\sim LZSP_{2}, \beta\sim S_{2}$\end{tabular}                 \\ 

\hline

\end{tabular}
\vspace{2mm}

\caption{Payoff matrix based on strategic adoptions.}
\label{tab1}
\end{center}
\end{table}

Here are the strategic possibilities arising from Tab.\ref{tab1}.

The case $i_{1} \sim s_{1} / i_{2} \sim s_{1}$ - (Social Optimum/Nash Equilibrium): The buyer and seller opt for an honest strategy - this is the best outcome possible that favours the well-being of each (seller and buyer) and their personal interests. In this case the sum of the individual utilities is the greatest.

The case $i_{1} \sim s_{2} / i_{2} \sim s_{1}$: The seller acts honestly but the buyer decides to adopt a dishonest behaviour - the seller is then paid as if the transaction had taken place correctly, while the buyer is penalized within the network considerably.

The case $i_{1} \sim s_{1} / i_{2} \sim s_{2}$: The buyer acts honestly but the seller decides to adopt a dishonest behaviour - the buyer is then paid as if the transaction had taken place correctly, while the seller is penalized within the network considerably.

The case $i_{1} \sim s_{2} / i_{2} \sim s_{2}$: Both parties decide to adopt a dishonest strategy and are then penalized within the network in a considerable way but still less than in the cases: $i_{1} \sim s_{2} / i_{2} \sim s_{1}$ and $i_{1} \sim s_{1} / i_{2} \sim s_{2}$ for the fraudulent party.

\subsubsection{LZSP token: A gateway to governance participation.}\label{governance}

As stated in the payoff functions definition, users who perform beneficial actions are awarded with LZSP tokens. The LZSP token stands for a gateway for its users towards the project's Decentralized Autonomous Organization (DAO), and consequently allows to initially vote (based on a ratio of 1 token = 1 vote) minor and major changes within the platform, the authenticity of the submitted certificates. 

This token is distributed during several actions defined and qualified as honest\footnote{Reference \cite{eco} for more details on defined reward cases and other LZSP distribution formulas.} throughout the realization of trade, is granted essentially according to the following formula:

\begin{equation}
\eta(\tau_{LZS}) = \varphi_{LZS} (1-\frac{\alpha'}{\iota'})\label{equ}
\end{equation}
\vspace{+0.15cm}

Where we assume that $\varphi_{LZS}$ is the total value - here in LZS but applicable to any other asset used to transact - of the transaction $\tau_{LZS}$ performed and/or concerned, $\alpha'$ has been fixed in the range $[1;100]$, and that $\eta$ is the value (in USD) of the LZSP distributed. 

The LZSP, bringing governance power, may be redeemed at a 1:1 ratio with the LZS token, the main project token with devoted liquidity (i.e., economic power).

\subsection{Decentralized Payment Refund Service}\label{insurance}

The protocol  ``Payment with Dispute Resolution'' (PwDR) allows honest victims of online payment fraud to prove their innocence and receive compensation for their financial losses \cite{insur}. PwDR is mainly based on efficient cryptographic primitives and smart contracts. It also involves a committee of auditors which processes victims' claims and decides where a victim should be reimbursed. It also preserves the privacy of the parties involved, e.g., users, victims and auditors. 

In our platform, to facilitate the reimbursement of victims, we propose to use the following idea. First, the victims prove their cases to the decentralized Kleros court \cite{lesaege2021long} (see Section \ref{conflict}). Then once the judgment is returned by the random jury and is considered fair by the judged parties, the winner of the dispute (seller or buyer) is reimbursed to the extent of his loss. If the judgment is considered unfair, it is possible for the wronged party to appeal.

As such, we allow an honest victim to access an amount equivalent to the relative prejudice/fraud, extracted from a collective liquidity pool that acts as an insurance, once their case has been ruled in their favor by a third-party jury.

\section{Conclusion}

The streetwear market is growing at an unprecedented rate. 
This important growth is further characterized by a concentration of several forms of fraud, often carried out on a large scale, which we intend to address through this paper.
We analyzed the most recurrent cases of fraud in P2P exchanges in the sneaker resale market and quantified their solvency levels. 
We find that counterfeit assets, payment scams, poor customer services, ineffective cancellations, unaccepted returns, hidden fees and delivery times are the major issues plaguing this growing industry today and are solvable.

To circumvent these issues, we propose :
(i) verification of assets submitted on the platform by a certified external verifier and the generation of an NFT linked to a genuine asset to track its ownership, attest to its authenticity and address counterfeiting issues;
(ii) a decentralized game-theoretic system, with no central entity involved, that provides economic incentives for users to behave honestly when selling/buying a P2P asset to prevent payment scams and delivery delays;
(iii) to support a decentralized dispute management protocol to address poor customer service, inefficient cancellations and unaccepted returns; and 
(iv) a transparent and autonomous blockchain-based trading system managed by smart contracts to address hidden fees. 
With these rules, we expect to improve trust, transparency and efficiency within the actual retail sneaker secondary market.

In future work, we will examine the practical implementation and potential limitations of our proposed mechanisms.
Factors such as the scalability of blockchain networks, the reliability of external authentication parties, and the effectiveness of decentralized dispute resolution mechanisms will be thoroughly evaluated and tested.
Education and user-friendly interfaces will also play a crucial role in facilitating the widespread adoption of the mechanisms.

\subsection{Acknowledgments}

We thank Dr. Abdelhakim Senhaji Hafid at the University of Montreal for his support and valuable improvements.

\bibliographystyle{splncs03}
\bibliography{Mybib.bib}

\begin{thebibliography}{10}
\providecommand{\url}[1]{\texttt{#1}}
\providecommand{\urlprefix}{URL }

\bibitem{insur}
Abadi, A., Murdoch, S.J.: Payment with dispute resolution: A protocol for
  reimbursing frauds victims. University College London  (2022),
  \url{https://eprint.iacr.org/2022/107}

\bibitem{alex}
Alexander, B., Bellandi, N.: Limited or limitless? exploring the potential of
  nfts on value creation in luxury fashion. The Journal of Design, Creative
  Process \& the Fashion Industry  (2022),
  \url{https://www.tandfonline.com/doi/abs/10.1080/17569370.2022.2118969}

\bibitem{prox}
Avrilionis, D., Hardjono, T.: From trade-only to zero-value nfts: The asset
  proxy nft paradigm in web3  (May 2022),
  \url{https://arxiv.org/pdf/2205.04899.pdf}

\bibitem{batra}
Batra, P., Singh, G.R., Gandhi, R.: Nft marketplace  (2023),
  \url{https://arxiv.org/abs/2304.10632}

\bibitem{pharma}
Bocek, T., Rodrigues, B.B., Strasser, T., Stiller, B.: Blockchains everywhere -
  a use-case of blockchains in the pharma supply-chain. 2017 IFIP/IEEE
  Symposium on Integrated Network and Service Management (IM)  (2017),
  \url{https://ieeexplore.ieee.org/abstract/document/7987376}

\bibitem{vote}
Canon, D., Charles, G.U., Foley, E., Hasen, R.: Restoring trust in the voting
  process. Election Law Journal 20(2)  (May 2021)

\bibitem{cassidy}
Cassidy, N.G.: The effect of scarcity types on consumer preference in the
  high-end sneaker market  (2018),
  \url{https://libres.uncg.edu/ir/asu/f/Cassidy_Nick\%20Spring\%202018\%20Thesis.pdf}

\bibitem{distrust}
Catulli, M.: What uncertainty?: Further insight into why consumers might be
  distrustful of product service systems  (July 2012),
  \url{https://www.researchgate.net/publication/235298357_What_uncertainty_Further_insight_into_why_consumers_might_be_distrustful_of_product_service_systems}

\bibitem{legit}
CheckCheck: Checkcheck website \url{https://getcheckcheck.com/en/}

\bibitem{choi}
Choi, Y., Lee, K.: Recent changes in consumer perception in sneaker resale
  market. International Journal of Costume and Fashion  (2021),
  \url{https://ijcf.ksc.or.kr/xml/29664/29664.pdf}

\bibitem{Paypalscam}
Community, P.: Security and fraud archives, elaborate sneaker scam please help
  \url{https://www.paypal-community.com/t5/Security-and-Fraud-Archives/Elaborate-sneaker-scam-please-help/td-p/2928591}

\bibitem{expen}
Edler, B.: Why sneakers are getting more expensive than ever. Complex  (March
  2015),
  \url{https://www.complex.com/sneakers/2015/03/why-sneakers-are-more-expensive}

\bibitem{adidas}
Effect, S.: Can you cancel adidas confirmed order?  (March 2023),
  \url{https://shoeeffect.com/can-you-cancel-adidas-confirmed-order/}

\bibitem{emma}
Emma, K.: The future of luxury fashion brands through nfts. Aalto University
  (2022), \url{https://aaltodoc.aalto.fi/handle/123456789/114089}

\bibitem{customerservice}
Forbes: Businesses lose \$75 billion due to poor customer service  (May 2018),
  \url{https://www.forbes.com/sites/shephyken/2018/05/17/businesses-lose-75-billion-due-to-poor-customer-service/}

\bibitem{tr}
Frignani, A., Bernard, J.: Innovative scan app as a means to overcome
  counterfeit in the sneaker market  (March 2022),
  \url{https://www.theseus.fi/bitstream/handle/10024/749736/Th\%C3\%A8se\%20Alan\%20frignani\%20Jules\%20Bernard.pdf?sequence=2}

\bibitem{GOATreview}
GOAT: Goat - comment review
  \url{https://www.trustpilot.com/review/goat.com?page=6&stars=1}

\bibitem{GOAT}
GOAT: Goat website \url{https://www.goat.com/}

\bibitem{commnet}
Han, Z., Niyato, D., Saad, W., Basar, T.: Game theory in wireless and
  communication networks: Theory, models, and applications  (January 2011)

\bibitem{attr}
Housley, R., Ford, W., Polk, W., , Solo, D.: Internet x.509 public key
  infrastructure certificate and crl profile  (January 1999),
  \url{http://tools.ietf.org/rfc/rfc2459.txt}

\bibitem{2ndmarket}
Hyun, S.J., Koh, B.: Benefiting from resellers: The impact of the online
  secondary sneaker market on the sneaker brand's profit and consumer surplus
  (2020), \url{https://aisel.aisnet.org/pacis2020/168/}

\bibitem{collab}
Iran, S., Schrader, U.: Collaborative fashion consumption and its environmental
  effects. Journal of Fashion Marketing and Management 21(4):00-00  (August
  2017),
  \url{https://www.researchgate.net/publication/319067796_Collaborative_fashion_consumption_and_its_environmental_effects}

\bibitem{joy}
Joy, A., Zhu, Y., Peña, C., Brouard, M.: Digital future of luxury brands:
  Metaverse, digital fashion, and non-fungible tokens. Special Issue: Luxury,
  Entrepreneurship, and Innovation, Part I  (2022),
  \url{https://onlinelibrary.wiley.com/doi/abs/10.1002/jsc.2502}

\bibitem{fincontr}
Karaivanov, A.K.: Blockchains, collateral and financial contracts. Project:
  Economics of blockchains, Simon Fraser University  (October 2019)

\bibitem{assetsneak}
Kernan, J., Chen, O., Zuber, K., Orr, J.: Sneakers as an alternative asset
  class  (July 2020),
  \url{https://www.cowen.com/insights/sneakers-as-an-alternative-asset-class-part-ii/}

\bibitem{pavel}
Kireyev, P.: Nft marketplace design and market intelligence. INSEAD Working
  Paper No. 2022/03/MKT  (2022),
  \url{https://papers.ssrn.com/sol3/papers.cfm?abstract_id=4002303}

\bibitem{trackmanagehealth}
Korpela, K., Novotny, P., Dubovitskaya, A., Dahlberg, T.: Blockchain design for
  digital supply chain integration. Artificial Intelligence for Sustainable and
  Resilient Production Systems (pp.90-98)  (August 2021)

\bibitem{kleros}
Lesaege, C., Ast, F.: Kleroterion, a decentralized court for the internet
  (June 2017),
  \url{https://www.researchgate.net/publication/318877800_Kleroterion_a_decentralized_court_for_the_Internet}

\bibitem{lesaege2021long}
Lesaege, C., George, W., Ast, F.: Kleros - long paper v2. 0.2  (July 2021)

\bibitem{ma}
Ma, K., Treiber, M.C.: Hedonic pricing in the sneaker resale market. Duke
  University  (2020),
  \url{https://sites.duke.edu/econhonors/files/2020/05/matreiber2020.pdf}

\bibitem{runrepeat}
McLoughlin, D.: Sneaker industry statistics  (October 2021),
  \url{https://runrepeat.com/sneaker-industry-stats}

\bibitem{sneakmarket}
Morency, C.: Special report: This is what the future of sneaker reselling looks
  like. Highsnobiety  (2019),
  \url{https://www.highsnobiety.com/p/sneaker-reselling-future/}

\bibitem{umer}
Muhammad, U.S., Vishwas, K.: Application of non-fungible tokens (nfts) and the
  intersection with fashion luxury industry  (2021),
  \url{https://www.politesi.polimi.it/handle/10589/182823}

\bibitem{lawsuit}
Pavlakos, L.: Stockx reaches settlement on class action lawsuit over hidden
  fees  (April 2023),
  \url{https://www.complex.com/sneakers/stockx-settles-class-action-suit-over-hidden-fees}

\bibitem{market}
PRNewswire: Global sneakers market stood at \$58 billion in 2018 and is
  projected to grow at a cagr of over 7\% during 2019-2024  (June 2019),
  https://www.prnewswire.com/news-releases/global-sneakers-market-stood-at--58-billion-in-2018-and-is-projected-to-grow-at-a-cagr-of-over-7-during-2019-2024-300861769.html

\bibitem{Nextgen}
Puri, N., Garg, V., Agrawal, R.: Blockchain technology applications for next
  generation. EAI/Springer Innovations in Communication and Computing book
  series (EAISICC)  (November 2021),
  \url{https://link.springer.com/chapter/10.1007/978-3-030-77637-4_4}

\bibitem{raditya}
Raditya, D., Erlin, N., Amanda, F., Hanafiah, N.: Predicting sneaker resale
  prices using machine learning  (2021),
  \url{https://www.sciencedirect.com/science/article/pii/S1877050921000430}

\bibitem{raman}
Raman, R., Raj, B.E.: The world of nfts (non-fungible tokens): The future of
  blockchain and asset ownership. Enabling Blockchain Technology for Secure
  Networking and Communications  (2021),
  \url{https://www.igi-global.com/chapter/the-world-of-nfts-non-fungible-tokens/280845}

\bibitem{driver}
Salpini, C.: What's driving retail's sneaker obsession? Retail Dive  (March
  2018),
  \url{https://www.retaildive.com/news/whats-driving-retails-sneaker-obsession/518625/}

\bibitem{sec}
Slaton, K., Pookulangara, S.: Collaborative consumption: An investigation into
  the secondary sneaker market. International IJC 46(10)  (May 2022),
  \url{https://www.researchgate.net/publication/352594333_Collaborative_Consumption_An_Investigation_into_the_Secondary_Sneaker_Market}

\bibitem{WashingtonPost}
Somasundaram, P.: As counterfeits rise, sneaker authenticators sniff out real
  from fake  (September 2022),
  \url{https://www.washingtonpost.com/business/2022/09/30/sneaker-authentication-ebay-stockx-nike/}

\bibitem{adidasscam}
SpamTitan: Don’t be fooled by this adidas phishing scam!  (February 2021),
  \url{https://www.spamtitan.com/blog/dont-be-fooled-by-this-adidas-phishing-scam/}

\bibitem{StockXreview}
StockX: Stockx - comment review
  \url{https://www.trustpilot.com/review/stockx.com}

\bibitem{StockX}
StockX: Stockx website \url{https://stockx.com/}

\bibitem{datastockx}
StockX: Stockx snapshot: The state of resale  (2020),
  \url{https://stockx.com/news/state-of-resale/}

\bibitem{2021-Half-Year-Fraud-Update}
{UK Finance}: 2021 half year fraud update (2021),
  \url{https://www.ukfinance.org.uk/system/files/Half-year-fraud-update-2021-FINAL.pdf}

\bibitem{white}
White, B., Mahanti, A., Passi, K.: Characterizing the opensea nft marketplace.
  WWW '22: Companion Proceedings of the Web Conference 2022  (2022),
  \url{https://dl.acm.org/doi/abs/10.1145/3487553.3524629}

\bibitem{eco}
Zeggari, M., Lambiotte, R.: Lyzis labs economic design  (2023)

\bibitem{lyz}
Zeggari, M., Lambiotte, R., Abadi, A., Axon, L., Kassab, M.: An efficient and
  decentralized blockchain-based commercial alternative (full version)
  (November 2022), \url{https://arxiv.org/pdf/2210.08372.pdf}

\bibitem{spec}
Zhang, Y.: Analysis of the causes, the influential aspects and the risk of
  sneaker speculation and suggestion. Proceedings of Business and Economic
  Studies  (June 2020),
  \url{https://www.researchgate.net/publication/348073592_Analysis_of_the_Causes_the_Influential_Aspects_and_the_Risk_of_Sneaker_Speculation_and_Suggestion}

\bibitem{zhu}
Zhu, C.: Dynamic double auctions: An analysis of secondary sneaker market and
  its future as an nft marketplace. Colby College  (2022),
  \url{https://digitalcommons.colby.edu/honorstheses/1378/}

\bibitem{NFTsneakmarket}
Zhu, C.: Nft sneaker marketplace design, testing, and challenges. Colby College
   (2022), \url{https://digitalcommons.colby.edu/honorstheses/1385/}

\end{thebibliography}

\appendix

\clearpage
\section{Appendix}\label{AppendixA}

This Appendix gives details on the market survey performed on the centralized sneaker resale marketplaces StockX \cite{StockX} and Goat \cite{GOAT}.

\subsection{Cross-referenced Analysis}

\begin{figure}[htpb!]
\centerline{\includegraphics[width=0.65\textwidth]{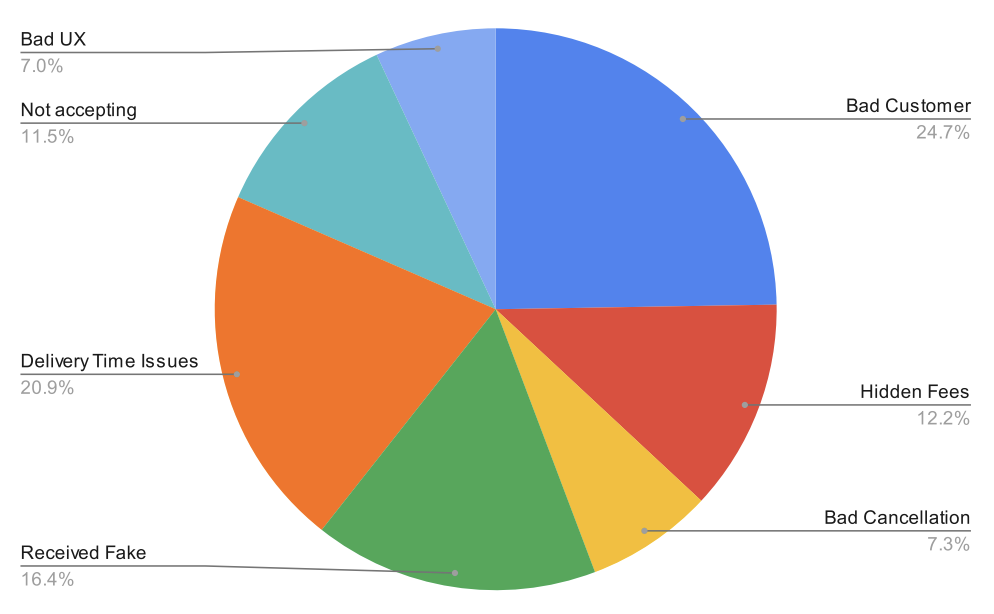}}
\caption{Cross-analysis of reviews with a set of n=287 to get a statistical significance with 90\% confidence level.}
\label{Fig5}
\end{figure}

\begin{table*}[htpb]
\begin{center}
\begin{tabular}{|l|l|}
\hline
\textbf{Type} & \textbf{Distribution} \\ \hline
Bad Customer Service & 71 \\ \hline
Hidden Fees & 35 \\ \hline
Bad Cancellation Policy & 21 \\ \hline
Received Fake & 47 \\ \hline
Delivery Time Issues & 60 \\ \hline
Not accepting Returns & 33 \\ \hline
Bad UX & 20 \\ \hline
\textbf{Total} & \textbf{287} \\ \hline
\end{tabular}
\vspace{2mm}
\caption{Overview of the distribution of the types of issues observed on StockX \cite{StockXreview} and GOAT \cite{GOATreview}.}
\label{tab3}
\end{center}
\end{table*}

\clearpage

\subsection{StockX Analysis}

\begin{table}[h!]
\begin{center}
\begin{tabular}{|l|l|l|}
\hline
Total Reviews & 45,000 & 100\% \\ \hline
1 Star Reviews & 12,150 & 27\% \\ \hline
\end{tabular}
\vspace{2mm}
\caption{Index of the number of 1 star reviews on the total number of StockX reviews \cite{StockXreview}.}
\label{tab6}
\end{center}
\end{table}

\begin{figure}[htpb!]
\centerline{\includegraphics[width=0.65\textwidth]{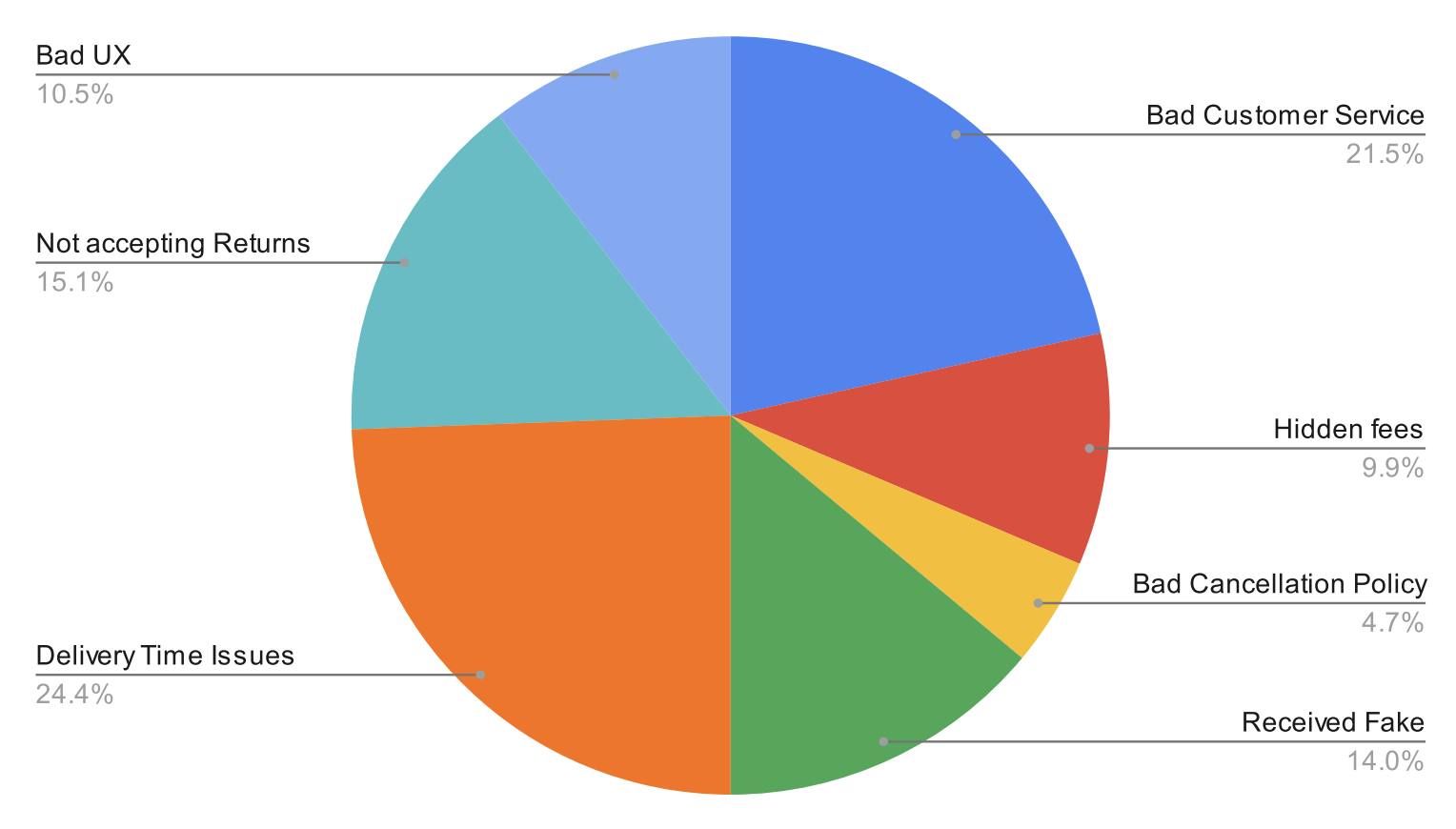}}
\caption{Analysis of StockX reviews \cite{StockXreview} with a set of n=172>150 to get a statistical significance with 90\% confidence level.}
\label{Fig3}
\end{figure}

\begin{table}[h!]
\begin{center}
\begin{tabular}{|l|l|}
\hline
\textbf{Type} & \textbf{Distribution} \\ \hline
Bad Customer Service & 37 \\ \hline
Hidden Fees & 17 \\ \hline
Bad Cancellation Policy & 8 \\ \hline
Received Fake & 24 \\ \hline
Delivery Time Issues & 42 \\ \hline
Not accepting Returns & 26 \\ \hline
Bad UX & 18 \\ \hline
\textbf{Total} & \textbf{172} \\ \hline
\end{tabular}
\vspace{2mm}
\caption{Overview of the distribution of the types of issues observed on StockX \cite{StockXreview}.}
\label{tab4}
\end{center}
\end{table}

\clearpage
\subsection{GOAT Analysis}

\begin{table}[h!]
\begin{center}
\begin{tabular}{|l|l|l|}
\hline
Total Reviews & 12,150 & 100\% \\ \hline
1 Star Reviews & 2,187 & 18\% \\ \hline
\end{tabular}
\vspace{2mm}
\caption{Index of the number of 1 star reviews on the total number of GOAT reviews \cite{GOATreview}.}
\label{tab7}
\end{center}
\end{table}

\begin{figure}[htpb!]
\centerline{\includegraphics[width=0.65\textwidth]{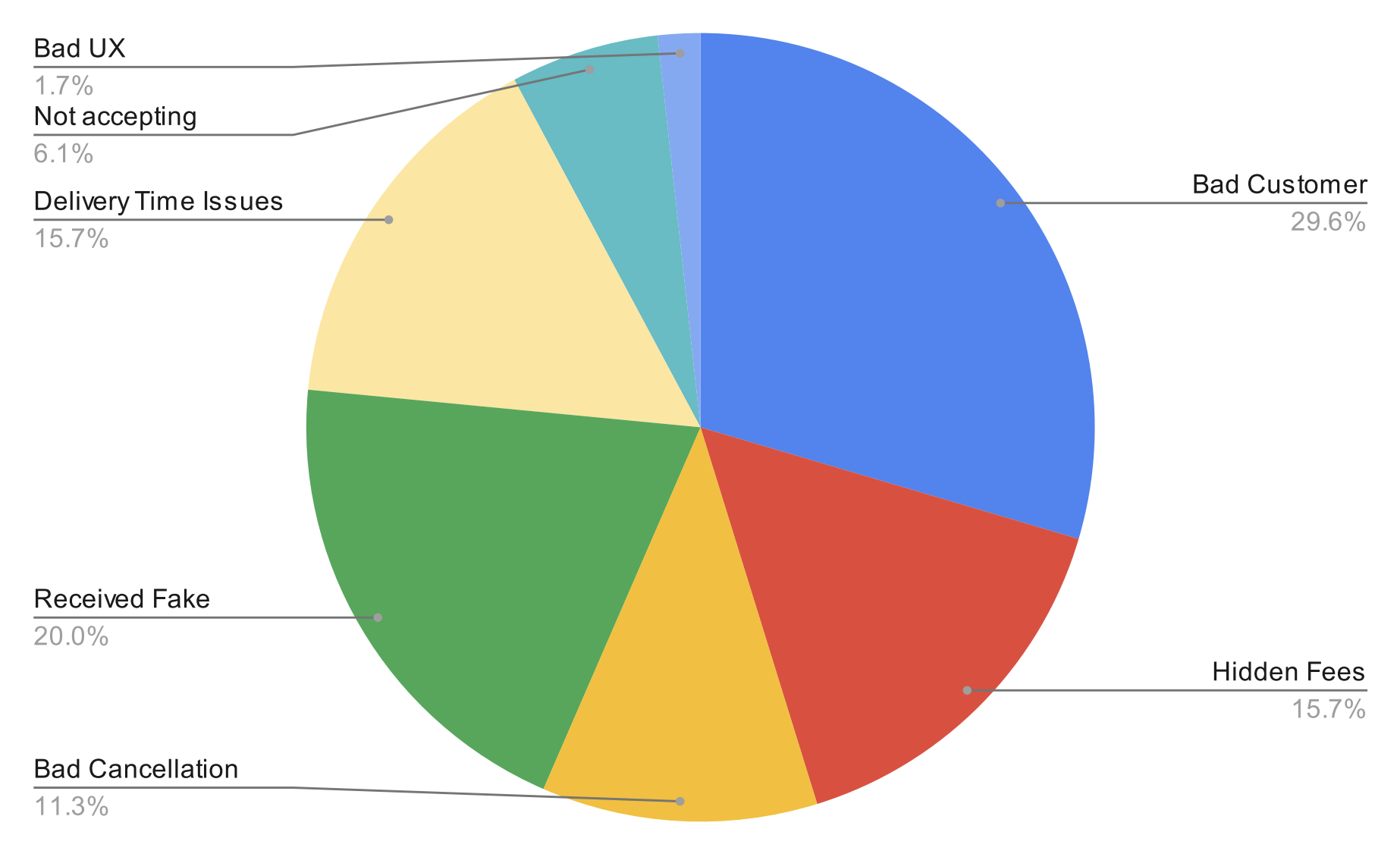}}
\caption{Analysis of GOAT reviews \cite{GOATreview} with a set of n=115 to get a statistical significance with 90\% confidence level.}
\label{Fig4}
\end{figure}

\begin{table}[h!]
\begin{center}
\begin{tabular}{|l|l|}
\hline
\textbf{Type} & \textbf{Distribution} \\ \hline
Bad Customer Service & 34 \\ \hline
Hidden Fees & 18 \\ \hline
Bad Cancellation Policy & 13 \\ \hline
Received Fake & 23 \\ \hline
Delivery Time Issues & 7 \\ \hline
Not accepting Returns & 18 \\ \hline
Bad UX & 2 \\ \hline
\textbf{Total} & \textbf{115} \\ \hline
\end{tabular}
\vspace{2mm}
\caption{Overview of the distribution of the types of issues observed on GOAT \cite{GOATreview}.}
\label{tab5}
\end{center}
\end{table}

\newpage 

\section{Appendix}\label{AppendixB}

This Appendix gives details on the technical design of the components integrated as solutions, i.e., the generated NFTs, the trading solution architecture and a descriptive of the symbols implied within the game theory application.

\begin{lstlisting}[language=Solidity, caption=Example of an NFT smart contract linked to a pair of physical sneakers., label=List1]
// SPDX-License-Identifier: MIT
pragma solidity ^0.8.0;

import "@openzeppelin/contracts/token/ERC721/ERC721.sol";
import "@openzeppelin/contracts/utils/Counters.sol";
 
contract NFT {
    uint256 id;
    string name;
    string imageUrl;
    string sneakerId;
    string location;
    string proofOfOwnership;
    address owner;

    constructor(uint256 _id, string memory _name, string memory _imageUrl, string memory _sneakerId, string memory _location, string memory _proofOfOwnership) public {
        id = _id;
        name = _name;
        imageUrl = _imageUrl;
        sneakerId = _sneakerId;
        location = _location;
        proofOfOwnership = _proofOfOwnership;
        owner = msg.sender;
    }

    function transferOwnership(address newOwner) public {
        require(msg.sender == owner, "Only the current owner can transfer ownership");
        owner = newOwner;
    }

    function getId() public view returns (uint256) {
        return id;
    }

    function getName() public view returns (string memory) {
        return name;
    }

    function getImageUrl() public view returns (string memory) {
        return imageUrl;
    }

    function getSneakerId() public view returns (string memory) {
        return sneakerId;
    }

    function getLocation() public view returns (string memory) {
        return location;
    }

    function getProofOfOwnership() public view returns (string memory) {
        return proofOfOwnership;
    }

    function getOwner() public view returns (address) {
        return owner;
    }
}
\end{lstlisting}

\begin{figure}[htpb!]
\centerline{\includegraphics[width=0.9\textwidth]{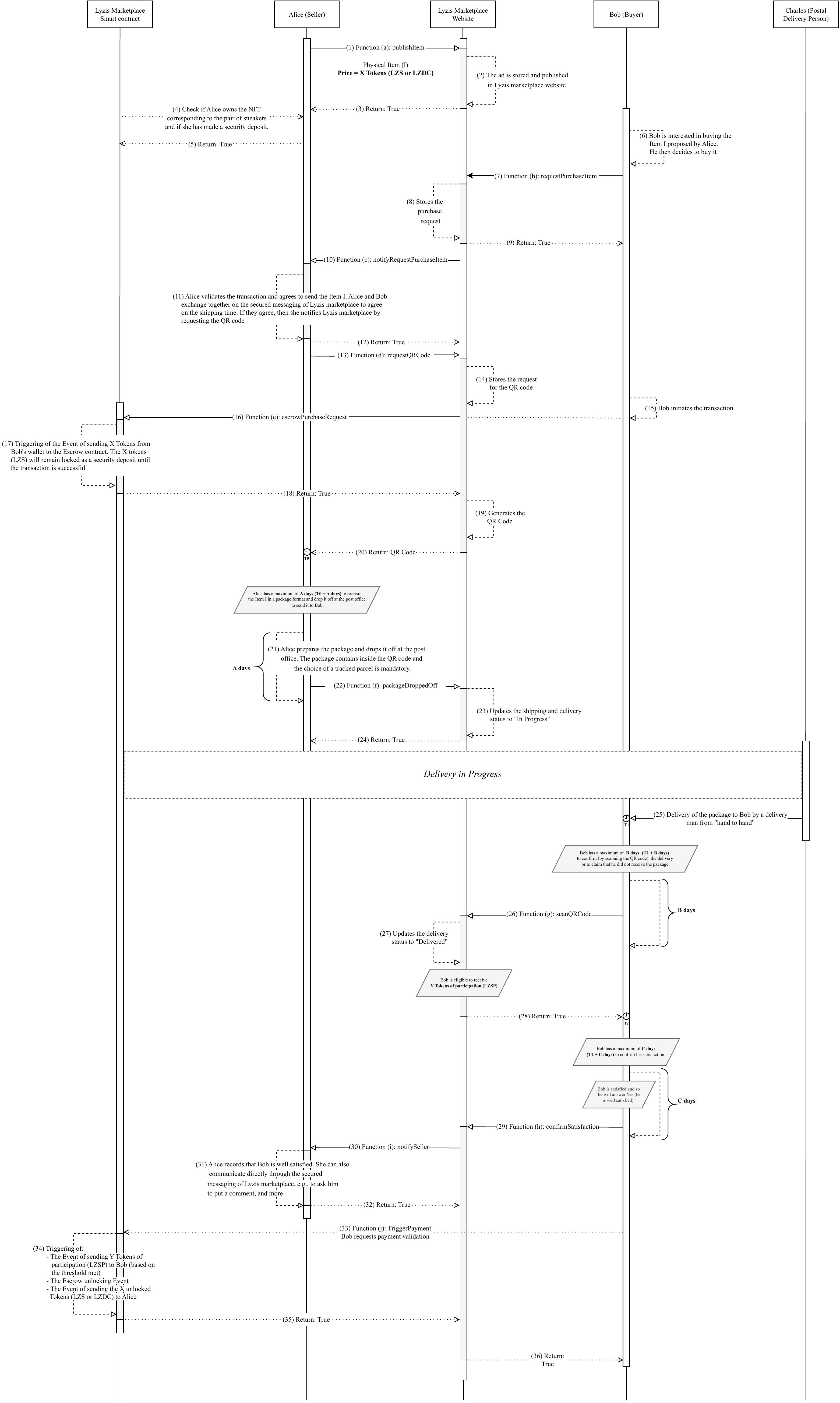}}
\caption{UML diagram of an example workflow of the proposed marketplace where Item I stands for a physical sneaker pair \cite{lyz}. We may integrate the \textit{'transferOfOwnership'} function of the NFT linked to Item I during steps (27) and (31) and a possible validation from (33).}
\label{Fig2}
\end{figure}

\begin{table}
\resizebox{\columnwidth}{!}{%
\begin{tabular}{|l|l|}
\hline
\textbf{Symbol} & \textbf{Definition} \\ \hline
\textit{$v_{1}$} & Monetary value involved in the transaction \\ \hline
\textit{$v_{2}$} & \begin{tabular}[c]{@{}l@{}}The tangible value relative to the physical asset\\ involved in the transaction (sneaker pair)\end{tabular} \\ \hline
\textit{$vo_{1}$} & A gain of the asset's value \\ \hline
\textit{$vo_{2}$} & A loss of the asset's value \\ \hline
\textit{$\alpha\sim LZSP_{1}$} & \begin{tabular}[c]{@{}l@{}}A gain of Lyzis participation/governance tokens \\ (LZSP)\end{tabular} \\ \hline
\textit{$\alpha\sim LZSP_{2}$} & \begin{tabular}[c]{@{}l@{}}A loss (and/or non-gain) of Lyzis participation/\\ governance tokens (LZSP)\end{tabular} \\ \hline
\textit{$\beta\sim S_{1}$} & \begin{tabular}[c]{@{}l@{}}A maintenance - by the seller - of the LZS tokens\\ initially stacked to validate the operation of his \\ account and the possibility for him to withdraw\\ the tokens following the exchange with the returns \\ granted based on the stacking duration\end{tabular} \\ \hline
\textit{$\beta\sim S_{2}$} & \begin{tabular}[c]{@{}l@{}}A loss - by the seller - of the LZS tokens initially \\ stacked to validate the operation of his account \\ and the impossibility for him to withdraw the tokens\end{tabular} \\ \hline
\textit{$\beta\sim S_{3}$} & \begin{tabular}[c]{@{}l@{}}A loss (non-gain) - by the buyer - of the LZS tokens\\ initially stacked by the seller to validate the \\ functioning of the account\end{tabular} \\ \hline
\textit{$\beta\sim S_{4}$} & \begin{tabular}[c]{@{}l@{}}A gain - by the buyer - of the LZS tokens initially \\ stacked by the seller to validate the functioning of\\ the account\end{tabular} \\ \hline
\textit{$\delta\sim R_{1}$} & \begin{tabular}[c]{@{}l@{}}A positive change in a user's reputation level on the \\ Lyzis Marketplace\end{tabular} \\ \hline
\textit{$\delta\sim R_{2}$} & \begin{tabular}[c]{@{}l@{}}A negative change in a user's reputation level on the \\ Lyzis Marketplace\end{tabular} \\ \hline
\end{tabular}%
}
\vspace{2mm}
\caption{Definition of payoff functions.}
\label{tab2}
\end{table}


\section{Appendix}\label{AppendixC}

\subsection{Non-tradable Value NFT: Free of Any Economic Valuation}

In order to best maintain consistency in identifying pairs of sneakers, the smart contract that runs the NFT manages the life cycle of the NFT by disabling any trading for payment feature \cite{prox}. Thus, we assume that the generated NFT is not intended for a secondary market where it can be economically untied from the asset it is designed to represent, allowing for a significant increase in consistency. We then also assume that the resulting NFTs don't represent any value as such and are then only used for the identification and tracking of an asset, being consequently free of any economic valuation, giving way to the total valuation put on the physical pair of sneakers.

\subsection{Hidden NFT Metadata}\label{C2}

Any information related to an NFT and considered sensitive, in this instance the serial number of the sneaker and/or personal informations such as location, must be kept inaccessible to any other party.
Hence, in order for the attributes of the NFT and the metadata of the sneaker pair to be selectively disclosed to authorized parties, the NFT must have only an opaque hash generated that solely identifies the serialized metadata related to the pair associated with the NFT.
Similarly, the asset's metadata must contain a reverse reference to the NFT identifier. Once the user with the appropriate identifying information requests the retrieval of the metadata information, he requests the resolution of the hashed metadata through an explicit request to the metadata repository \cite{prox}. This system is similar and can then leverage zero knowledge proofs (ZKP). As a result, it is no longer possible for any user other than the authorized users in possession of the hash to extract sensitive information about a pair of sneakers by merely reading the NFT information on-chain, except for basic information/attributes.
\end{document}